%% file: hadron2011.tex
\documentclass[a4paper,11pt]{article}

\usepackage{contribution}

\input{econfmacros}
\input{contributionmacros}

\begin{document}

\input{contribution}

\end{document}

%% file: econfmacros.tex
\newcommand{\weblink}[2][]{%
    \ifthenelse{\equal{#1}{}}%
    {\textnormal{\url{#2}}}%
    {\textnormal{\href{#2}{#1}}}%
}

\newcommand{\babar}{\mbox{\slshape B\kern-0.1em{\smaller A}\kern-0.1em B\kern-0.1em{\smaller A\kern-0.2em R} }}

\def\beq{\begin{equation}}
\def\eeq#1{\label{#1}\end{equation}}
\def\eeqn{\end{equation}}

\def\beqa{\begin{eqnarray}}
\def\eeqa#1{\label{#1}\end{eqnarray}}
\def\eeqan{\end{eqnarray}}

\let\bar=\overbar

\def\Dslash{\not{\hbox{\kern-4pt $D$}}}
\def\dslash{\not{\hbox{\kern-2pt $\del$}}}

\def\msb{{\bar{\ssstyle M \kern -1pt S}}}

%% file: contributionmacros.tex
\newcommand{\contribution}[7][]{%
  \clearpage
  \thispagestyle{plain}
  \ifthenelse{\equal{#1}{}}
  {\hypersetup{pdftitle={#2}}}
  {\hypersetup{pdftitle={#1}}}
  \hypersetup{pdfauthor={{#3} {#4}}}
  {\centering\normalfont\LARGE\bfseries\sffamily #2 \par\nobreak}
  \lhead{}
  \chead{%
    \textit{\footnotesize XIV International Conference on Hadron Spectroscopy
      (\weblink[\textit{hadron2011}]{http://www.hadron2011.de}), 13-17 June 2011, Munich, Germany}%
  }
  \rhead{}
  \bigskip
  \begin{center}
    {#3} {#4}\ifthenelse{\equal{#6}{}}{}{\footnote{\weblink[#6]{mailto:#6}}}
    \ifthenelse{\equal{#7}{}}{}{#7} \\
    \textit{#5}
  \end{center}
  \bigskip
}

\renewcommand{\abstract}[1]{%
  \begin{center}
    \begin{minipage}{0.85\textwidth}
      \begin{footnotesize}
        #1
      \end{footnotesize}
    \end{minipage}
  \end{center}
  \bigskip
}

%% file: contribution.tex
%
%
%
%
%
{  


%

\contribution  
{Charmonium \& Charmonium-like States with \babar}  
{Valentina}{Santoro}  
{INFN Ferrara , via Saragat 1, Ferrara, 44122, ITALY}  
{santoro@fe.infn.it}  
{on behalf of the \babar Collaboration}  
%

\abstract{%
  We review recent charmonium and charmonium-like state from the  \babar B-factory. A particular focus is given  to the observation of the decay 
  $X(3872) \to J/\psi~\omega$ and to recent $\eta_{c}(1S)$ and $\eta_{c}(2S)$ results. The Observation of the $\chi_{c2}(2P)$ state will be also discussed. }
%

\section{Introduction}

The charmonium spectrum consists of eight narrow states below the open charm threshold (3.73~GeV) and several tens of states above that. Below the threshold almost all states are well established. On the other hand, very little is known at higher energy. Only one state has been positively identified as a charmonium $D$ state, the $\psi (3770)$, in addition there are several new ``Charmonium-like" states that are very difficult to accommodate in the conventional charmonium spectrum. 

\section{Evidence for the decay $X(3872) \to J/\psi~\omega$ }
The X(3872) it has been observed by several experiments in different decay modes and two production channels. Soon after its discovery \cite{Choi:2003ue} a great deal of effort has been expanded to understand the nature of the X(3872), especially its spin-parity assignment; so far, only $J^{PC}=1^{++}$ or $2^{-+}$ can be assigned to the X(3872). In a previous \babar analysis  \cite{Aubert:2007vj} of the decay $B\to J/\psi\omega K$ there was clear signal for the Y(3940) as reported by Belle  \cite{Abe:2004zs} but there was no evidence for the X(3872). In this analysis were required for the $\omega$ candidates $0.765~GeV/c^{2} \leq m_{3\pi}\leq 0.7965$ $GeV/c^{2}$. In a more recent \babar analysis  \cite{arafat} the same decay mode has been studied using a slightly larger dataset and a lower $\omega$ mass limit: $0.74~GeV/c^{2} \leq m_{3\pi}\leq 0.7965$ $GeV/c^{2}$. All other selection criteria are the same as in the previous analysis. The $J/\psi~\omega$ mass distribution after background subtraction is shown in Fig.  \ref{fig:magnet}. There is clear signal for the $Y(3940)$, and evidence for the $X(3872)$. These signals are present in both $B^{+}$ and $B^{0}$ samples. The $m_{J/\psi\omega}$ distributions are fitted simultaneously with a function with three components: a Gaussian function for the X(3872), a relativistic $S$-wave Breit-Wigner for the Y(3940) and a broad Gaussian function multiplied by $m_{J/\psi\omega}$ for the nonresonant contribution. From the fit the following parameters are obtained: $m_{X(3872)}=3873.0^{+1.8}_{-1.6}(stat)\pm1.3(syst)~MeV/c^{2}$, $m_{Y(3940)}=3919.1^{+3.8}_{-3.4}(stat)\pm2.0(syst)~MeV/c^{2}$ and $\Gamma_{Y(3940)}=31^{+10}_{-8}\pm5~MeV$.
To investigate the parity of the X(3872) events with $3.8625~GeV/c^{2} \leq m_{J/\psi \omega} \leq 3.8825~GeV/c^{2}$ are selected. For these events the $m_{3\pi}$ is shown on Fig.~\ref{fig:m3pi} and is compared with the Monte Carlo simulation for different spin assignment. The $P$-wave assignment is favored ($\chi^{2}=3.53/5$) over the $S$-wave ($\chi^{2}=1017/5$) hence $J^{P}=2^{-}$ is favored over $J^{P}=1^{+}$.

\begin{figure}[htb]
  \begin{center}
    \includegraphics[width=0.41\textwidth]{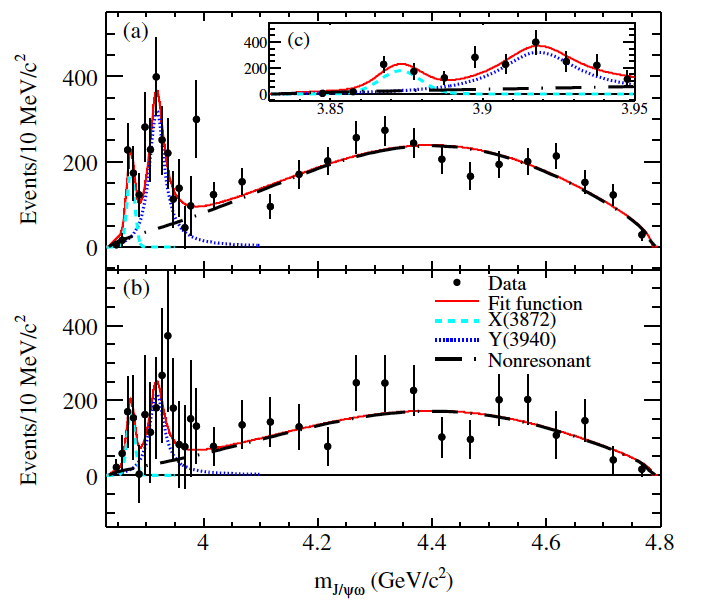}
    \caption{The $J/\psi\omega$ mass distribution for (a) $B^{+}\to J/\psi \omega K^{+}$ and (b) $B^{0}\to J/\psi\omega K^{0}_{s}$ decays; (c) shows the region $m_{J/\psi\omega} < 3.95~GeV/c^{2}$ of (a). The curves show the fit results and the individual fit contributions.}
    \label{fig:magnet}
  \end{center}
\end{figure}
\begin{figure}[htb]
  \begin{center}

        \includegraphics[width=0.31\textwidth]{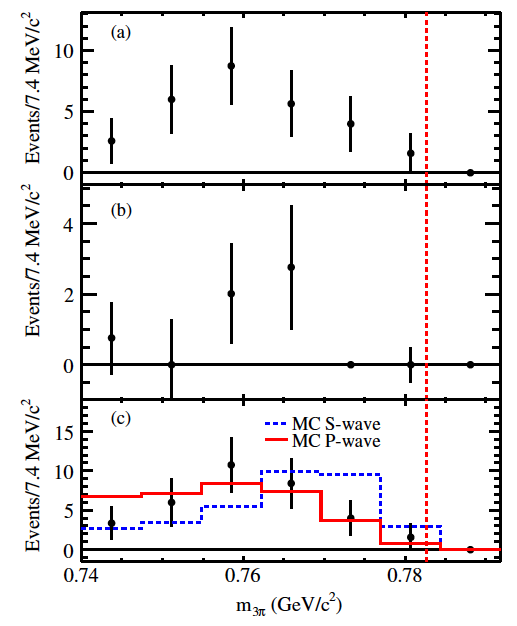}
    \caption{The $m_{3\pi}$ distribution for events $3.8625~GeV/c^{2} \leq m_{J/\psi \omega} \leq 3.8825~GeV/c^{2}$ for (a) $B^{+}$, (b) $B^{0}$, and (c) combined. The vertical line shows the $\omega$ nominal mass. In (c), the solid (dashed) histogram shows the $P$-wave ($S$-wave) Monte Carlo events normalized to the number of data events.}
    \label{fig:m3pi}
  \end{center}
\end{figure}

\section{Observation of the $\chi_{c2}(2P)$ meson in the reaction $\gamma\gamma \to D\bar{D}$}
In 2006 the Belle collaboration observed the Z(3930) \cite{ue} in $\gamma \gamma$ production of the $D\bar{D}$ system and this is considered a strong candidate for
the $\chi_{c2}(2P)$ state \cite{ue}. A recent \babar  analysis  \cite{chi2p} confirmed this state, in the same production mechanism using a data sample of 384 $fb^{-1}$. After all the selection criteria applied the efficiency corrected $D\bar{D}$ mass distribution is shown in Fig. \ref{fig:chi2p}. There is clear signal for the Z(3930) with a significance of 5.8 $\sigma$. An unbinned maximimum
likelihood fit is performed to obtain the following parameters: $m_{Z(3930)}=3926.7\pm2.7\pm1.1~MeV/c^{2}$ and $\Gamma_{Z(3930)}=21.3\pm6.8\pm 3.6~MeV$. \\
A decay angular analysis provides evidence that the Z(3930) is a tensor state with positive spin parity and C-parity $(J^{PC}=2^{++})$ in agreement with the $\chi_{c2}(2P)$ interpretation. The value of the partial width $\Gamma_{\gamma \gamma} \times BF(Z(3930)  \rightarrow D \bar{D})$ is found to be $(0.241 \pm 0.054 \pm 0.043)$ keV. All these measurements are in good agreement with the previous Belle measurements .

\begin{figure}[htb]
  \begin{center}
    \includegraphics[width=0.37\textwidth]{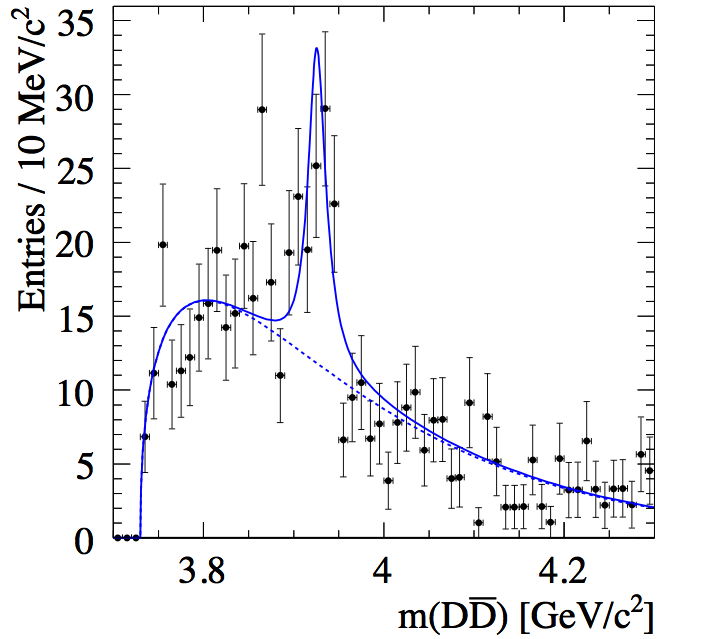}
    \caption{Efficiency corrected $D\bar{D}$ mass distribution with standard fit. The dotted line shows the background lineshape.}
    \label{fig:chi2p}
  \end{center}
\end{figure}

\section{Observation of the $\eta_{c}(1S)$ and $\eta_{c}(2S)$ decays to $K^{+}K^{-}\pi^{+}\pi^{-}\pi^{0}$ }
\babar studied using 519.2 $fb^{-1}$ collected at the $\Upsilon(4S)$, $\Upsilon(3S)$ and $\Upsilon(2S)$ the processes $e^{+}e^{-}\to e^{+}e^{-} \gamma \gamma$ $\to e^{+}e^{-} f$ where $f$ denotes $K^{+}K^{-}\pi^{+}\pi^{-}\pi^{0}$ or $K_{S}^{0}K^{\pm}\pi^{\mp}$ final states \cite{biassoni}. The allowed $J^{PC}$ values of the initial state are $0^{\pm+}, 2^{\pm+}, 4^{\pm+}, ...; 3^{++}, 5^{++}$. Angular momentum, parity conservation, and charge conjugation invariance, then imply that these quantum numbers apply to the final states $f$ also, except that the
$K_{S}^{0}K^{\pm}\pi^{\mp}$ state cannot have $J^{P} = 0^{+}$. The $K_{S}^{0}K^{\pm}\pi^{\mp}$ and $K^{+}K^{-}\pi^{+}\pi^{-}\pi^{0}$ mass spectra are shown on Fig. \ref{fig:eta}.
 There are signals at the position of the $\eta_{c}(1S)$, $J/\psi$, $\chi_{c0}(1P)$, $\chi_{c2}(1P)$ and $\eta_{c}(2S)$ states. The mass and the width of the $\eta_{c}(1S)$ and $\eta_{c}(2S)$, extracted using a binned extended maximum likelihood fit, are respectively $m_{\eta_{c}(1S)}=2982.5 \pm 0.4 \pm 1.4 ~\mathrm{MeV/c^{2}}$, $m_{\eta_{c}(2S)}=3638.5 \pm 1.5 \pm 0.8 ~\mathrm{MeV/c^{2}}$ and $\Gamma_{\eta_{c}(1S)}=32.1\pm  1.1 \pm 1.3 ~\mathrm{MeV}$, $\Gamma_{\eta_{c}(2S)}=13.4 \pm 4.6 \pm 3.2~\mathrm{MeV}$.

\begin{figure}[htb!]
  \begin{center}
    \includegraphics[width=0.4\textwidth]{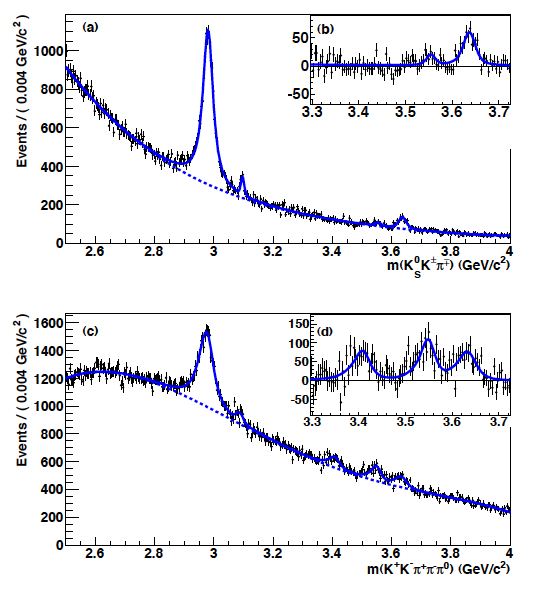}
    \caption{Fit to the $K_{S}^{0}K^{\pm}\pi^{\mp}$ (a) and $K^{+}K^{-}\pi^{+}\pi^{-}\pi^{0}$ (c) mass spectra. The solid curves represent the total fit functions and the dashed curves show the combinatorial background contributions. The background-subtracted distributions are shown in (b) and (d).}
    \label{fig:eta}
  \end{center}
\end{figure}


%

}  
